\documentclass[prl,twocolumn,superscriptaddress,notitlepage,9py]{revtex4-1}
\usepackage[utf8]{inputenc}
\usepackage{natbib}
\usepackage{graphicx}
\usepackage{color}
\usepackage[left=1.5cm, right=1.5cm, top=1.785cm, bottom=2.0cm]{geometry}
\usepackage{amsmath}
\usepackage{floatrow}

\begin{document}

\title{\LARGE Relationships among structure, memory, and flow in sheared disordered materials}

\author{{\Large K.L. Galloway$^{1}$, E.G. Teich$^{2}$, X-g  Ma $^{3}$, Ch. Kammer$^{1}$, I.R. Graham$^{3}$, N.C. Keim$^{4}$, C. Reina$^1$, D.J. Jerolmack$^{5,1}$, A. G. Yodh$^{3}$, P.E. Arratia$^{*,1}$ \\}
{\large \small $^{1}$Department of Mechanical Engineering and Applied Mechanics, University of Pennsylvania \\ $^{2}$Department of Bioengineering, University of Pennsylvania \\ 
$^{3}$Department of Physics and Astronomy, University of Pennsylvania \\ 
$^{4}$Department of Physics, Pennsylvania State University \\ 
$^{5}$Department of Earth and Environmental Science, University of Pennsylvania \\ 
}}

\date{\today}

\begin{abstract}
    \normalsize{\textbf{ A fundamental challenge for disordered solids is predicting macroscopic yield from the microscopic arrangements of constituent particles. Yield is accompanied by a sudden and large increase in energy dissipation due to the onset of plastic rearrangements. This suggests that one path to understanding bulk rheology is to map particle configurations to their mode of deformation. Here, we perform laboratory experiments and numerical simulations that are designed to do just that: 2D dense colloidal systems are subjected to oscillatory shear, and particle trajectories and bulk rheology are measured. We quantify particle microstructure using excess entropy. Results reveal a direct relation between excess entropy and energy dissipation, that is insensitive to the nature of interactions among particles. We use this relation to build a physically-informed model that connects rheology to microstructure. Our findings suggest a framework for tailoring the rheological response of disordered materials by tuning microstructural properties.}}
\end{abstract}

\maketitle

\section{Introduction}

Disordered solids are ubiquitous. They are found, for example, in our foods as pastes and gels \cite{nagel17}, and amidst our homes in the form of concrete \cite{ioan16} and mud \cite{jerol19,nie20}. Frustratingly, these materials can experience sudden mechanical failure, such as the collapse of soil during rapid mudslides. Indeed, when sufficiently stressed, all disordered materials exhibit a swift decrease in ability to support load. In the vicinity of this “yield” transition, the solid material shifts from a state wherein energy is stored via internal elastic forces, to a state in which energy is dissipated via irreversible plastic rearrangements \cite{larson99,guaz18,chen10,butt17,pham08}. Microscopic spatiotemporal features are associated with this yield transition and affect macroscopic material responses such as ductile versus brittle behavior. Unfortunately, in contrast to the case for crystalline materials, our ability to predict and control yield in disordered solids based on their constituents and their interactions is still limited \cite{guaz18,cipel20}. To build such microstructural models, we need to identify key microscopic metrics relevant to plasticity in disordered materials \cite{rich20}. Recently, excess entropy has been explored for this purpose \cite{gallo20b,inge17, bonn20}. In equilibrium systems, excess entropy has been utilized to connect viscosity with interparticle structure \cite{dzu96,rose99,dyre18,ma19}. Recently in far-from-equilibrium systems, excess entropy scaling has been shown to facilitate a relationship between microscopic structure and dynamics \cite{inge17, bonn20, gallo20b}. Thus, excess entropy offers an untapped signature for plasticity and a potential tool for modeling the mechanical response of disordered solids.  

The study of rheology and particle dynamics in disordered systems has a venerable history \cite{argon79}. As a result of this research, theories \cite{rich20} have proliferated in recent decades. Two of the most successful are Mode Coupling Theory, wherein the interplay of dynamical modes causes the emergence of rearrangements \cite{sieb09,fuchs10}, and Shear Transformation Zone theory, which posits that local configurations determine where rearrangements occur \cite{argon79,falk11,falk98,slott12}. More recently, structural signatures for rearrangement have been revealed by machine learning approaches \cite{cubuk17, cubuk15, bapst20, ma19a}, by study of low-frequency excitations \cite{ chen10a,yunk11,chen11,seth11,khab20,xu07}, and via local yield stress \cite{patin16} and near-neighbor cage dynamics \cite{maes17}. Despite their usefulness, difficulties remain in applying these theories to experiments because of the need for fitting parameters \cite{bouch11, sieb09} and the use of empirical relations\cite{maes17} that are difficult to measure. Moreover, these theories typically do not account for history-dependent behavior such as material memory, which is necessary to understand plasticity. 

Generally, disordered materials contain memories, i.e., microscopic signatures related to how the material has been processed \cite{keim11,keim19,muk19,pash19,fioc14,schwen20,mung19}. Memory of a previous shearing direction, for example, can be encoded into a material's response; once a material is sheared sufficiently in a given direction, continued shear in that direction requires more force than in the opposite direction \cite{gada80,keim19}. In jammed systems, recent experiments and simulations have studied formation of directional memory at low strain amplitudes, both below and near the yield transition; far above yield, memories are erased \cite{keim13a,paul14,teich20}. These observations, in turn, raise important new questions: do memories require elastic storage? Is plasticity synonymous with erasure? How do these phenomena manifest during yield, e.g., in storage and loss moduli?    

In this contribution, we utilize excess entropy to quantify material memory and construct a microstructural model for disordered-material response and energy dissipation. Experiments and simulations show that three non-dimensional parameters govern the connections between microstructure and bulk rheology: packing density, a normalized (non-dimensional) form of the imposed stress, and an excess entropy (microstructure-related) ratio that quantifies the material's ability to retain information about its initial state. Our results confirm that memory is stored elastically and lost plastically, and show how yield and the ductile/brittle response emerge from knowledge about particle configurations at the microscopic scale. 

\section{Results}

\begin{figure*}
\caption{\label{fig:1} \textbf{ Overview of structure, dynamics, and response.} \small{We characterize the disordered solid bulk response to cyclic stress from evolving configurations of individual constituent particles. (a) Image of $\sim$40,000 particles. Part of the raw image is shown (left). The scale bar is 200$\mu m$. Detected particle positions are also shown (right). For illustration, color indicates $D^2_{min,C}$, which quantifies the degree to which a particle has followed a non-affine returning trajectory (blue), or a non-affine escaping trajectory (red). The particles in this image are experiencing yield ($\gamma_0 \sim 15.7 \%$). (b) Quantification of the fractions of escaping and returning events versus total strain amplitude. Returning events rapidly increase near the yield point ($\gamma_0 \sim 3.0 \%$). (c) The number of particles, $Z(r)$ within a radius, $r$ of a reference particle. The radius is expressed in units of $a$, the average distance between neighboring particles. Vertical dashed lines indicate the limit of the first shell of neighboring particles. Inset: radial distribution function, $g(r)$. (d) The measured strain of the material versus the imposed stress throughout a cycle. Both stress and strain are averaged stroboscopically over 25 cycles. The different ellipses correspond to separate runs at different imposed stress amplitudes. Here, the area enclosed is a result of the lag between stress and strain, which in turn quantifies the energy dissipated from the material.} }
{\includegraphics{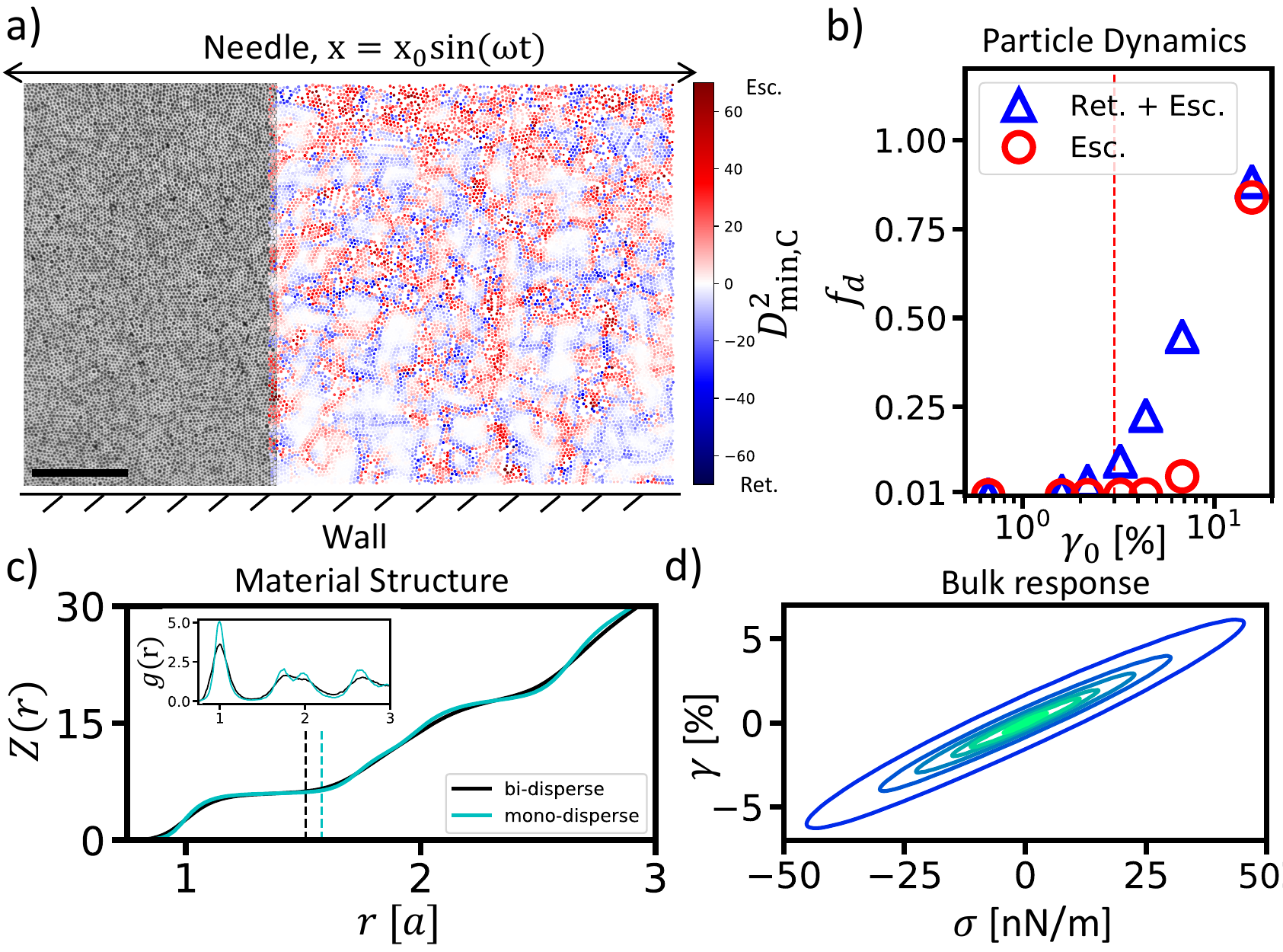}}
\end{figure*}

The experiments investigate disordered solids. The solids are colloidal monolayers of athermal, spherical particles ($\sim$ 40,000) adsorbed at an oil-water interface (Fig.~\ref{fig:1}a). The charged particle surfaces generate a dipole-dipole repulsion between particles. This repulsion is strong enough to jam the entire material, arresting particle motions. To probe the effects of disorder, we study both mono-disperse and bi-disperse spherical particle systems with diameters of 5.6 $\mu$m and 4.1 $\mu$m-5.6 $\mu$m, respectively. In the bi-disperse system, crystalline domains tend to be much smaller (See Supplementary Materials). We impose many cycles of sinusoidal stress on these samples using a custom-made interfacial stress rheometer \cite{keim13b}that permits measurement of the bulk response of the colloidal monolayer while simultaneously recording trajectories of individual particles (see Methods). Cyclic stress is quasi-static, insofar as the time scale for a completion of a rearrangement ($\sim$0.5s) is much shorter than the shortest driving period (5s) or largest inverse strain rate (20s). 

We investigate particle rearrangements by identifying non-affine deformations within each particle's neighborhood \cite{falk98,keim13b}. The degree of non-affinity is quantified by the mean-squared displacement after subtracting the best fit affine transformation, $D^2_{min}$ (see references \cite{falk98,keim13b} for more information). Within cyclically sheared disordered materials, two types of non-affine events occur (Fig.~\ref{fig:1}a): those wherein particles return to their original position at the end of a strain cycle but along different paths, and those wherein particles escape their nearest neighbors and do not return \cite{keim14,lund08,mobius14,regev13,gallo20a}. For visualization we define $D^2_{min,C} \equiv \pm \sqrt{(D^2_{min,R})^2+( D^2_{min,E})^2}$, where the subscripts refer to returning (R) and escaping (E) events, respectively, and sign corresponds to the greater $D^2_{min}$. Both types of events dissipate energy \cite{lund08,mobius14,keim14,regev13,regev14,gallo20a}. Returning non-affine events are known to emerge near the yield point when elasticity begins to diminish and plasticity starts to increase \cite{keim13b,butt17}; escaping events arise well beyond yield   \cite{keim13b} (Fig.~\ref{fig:1}b). The fraction of particles undergoing non-affine events is $f_d$. By following the rearrangements, we develop understanding about trajectory dynamics within the microstructure, and we take steps towards our ultimate goal to relate microstructure to rheology. 

To quantify structure, we characterize the inter-particle forces and particle configurations using the radial distribution function, $g(r)$.  Since the material is jammed, the motion of each particle is arrested by its neighbors \cite{hecke09,liu98,behr18,liu10,das20,hax12}. This caging, and escape thereof, provides another lens for the non-affine motions mentioned above; when enough particles pass each other via small changes in the structure of their surrounding cage, the material yields \cite{falk98}. For quantitative analysis, we compute $F^*$,  the sum of the magnitudes of inter-particle forces acting on the average particle. Specifically: $F^* = 2\pi\rho \int_0^{r_{N}}{ (-\frac{\partial u}{\partial r}) g(r) } rdr$; here $\rho$ is the number density of particles, $r_{N}$ is an upper cutoff distance below which nearest neighbor particles are found, $u$ is the pair potential function between any two particles, $\frac{\partial u}{\partial r}$ is the force acting between any two particles, and $g(r)$ is the sample radial distribution function as a function of separation $r$ (Fig.~\ref{fig:1}c; Methods). To determine $r_N$, we use the coordination number as a function of radial distance, $Z(r)$ (Fig.~\ref{fig:1}c). $Z(r)$ is derived from $g(r)$ and has been studied \cite{hecke09} and recently used \cite{maes17} to characterize particle interactions and their effect on bulk materials.  In our systems, neighbor shells are well defined by broad peaks in $g(r)$ separated by troughs (Fig.~\ref{fig:1}c-inset). The extent of the nearest neighbor shell is defined as the radius at which $Z(r)$ begins to increase rapidly for a second time (Fig.~\ref{fig:1}c-main). 

We quantify disorder using excess entropy \cite{rose99}, the difference between the system's entropy and that of its ideal gas analogue (identical pressure, temperature, etc.). The two-body approximation of excess entropy, $s_2$, is calculated from $g(r)$ using a formula given in the methods section (Eq.~\ref{excess_entropy}).   We calculate $s_2$ at discrete time points to characterize its variation within each shear cycle (more below). Since our systems are jammed, we interpret the below-yield system $s_2$ as `frozen in' excess entropy. 

We seek to relate these microstructural parameters to bulk rheological properties. Recall that as the yield transition is approached from below, the strain will begin to lag behind the oscillatory imposed stress by a phase angle, $\delta$. If $\delta=0[rad]$, then the material is fully elastic. If $\delta=\pi/2[rad]$, then the material is fully viscous. In between, the material exhibits both elasticity and plasticity; the phase angle lag quantifies dissipation (Fig.~\ref{fig:1}d) and encodes the ratio of the loss (plasticity) and storage (elasticity) moduli, $G^{\prime\prime}/G^{\prime} = tan(\delta)$. We will show how $G^{\prime\prime}/G^{\prime}$ is related to the microstructural and dynamical quantities described above ($s_2$, $F^*$, $f_d$). 

\begin{figure*}
\caption{ \label{fig:2} \textbf{ Memory within microstructure.} \small{Microstructural anisotropy reveals signatures of memory. Below yield, anisotropic orientation remains unchanged regardless of shear direction. Orientation quantifies stored memory. Above yield, anisotropic orientation reverses freely to match the direction of shear, indicating a loss of memory. (a) Radial distribution function, g(x,y,t) at a time corresponding to one quarter of the way through a shearing cycle. We fit an ellipse to the first neighbor ring. This ellipse stretches and reorients over time indicating changes in structural anisotropy of the sample. Two elliptic fits are shown at two times, t=1.25 (\textbf{---}) and 1.75 [cycles] (\textbf{- - -}). (b) Orientation of the sample microstructure over time as a function of strain amplitude. With increasing strain amplitude, the microstructure reorients to match the stretching axis. It first reorients completely at the yield point ($3.2\%$). (c) Elongation quantified by the ratio of ellipse major and minor axis lengths ($m/n$) over time. Below yield, elongation oscillates directly with the strain; above yield, elongation oscillates with twice the frequency of strain perturbation. In b \& c data are averaged strobscopically over 25 cycles.} }
{\includegraphics{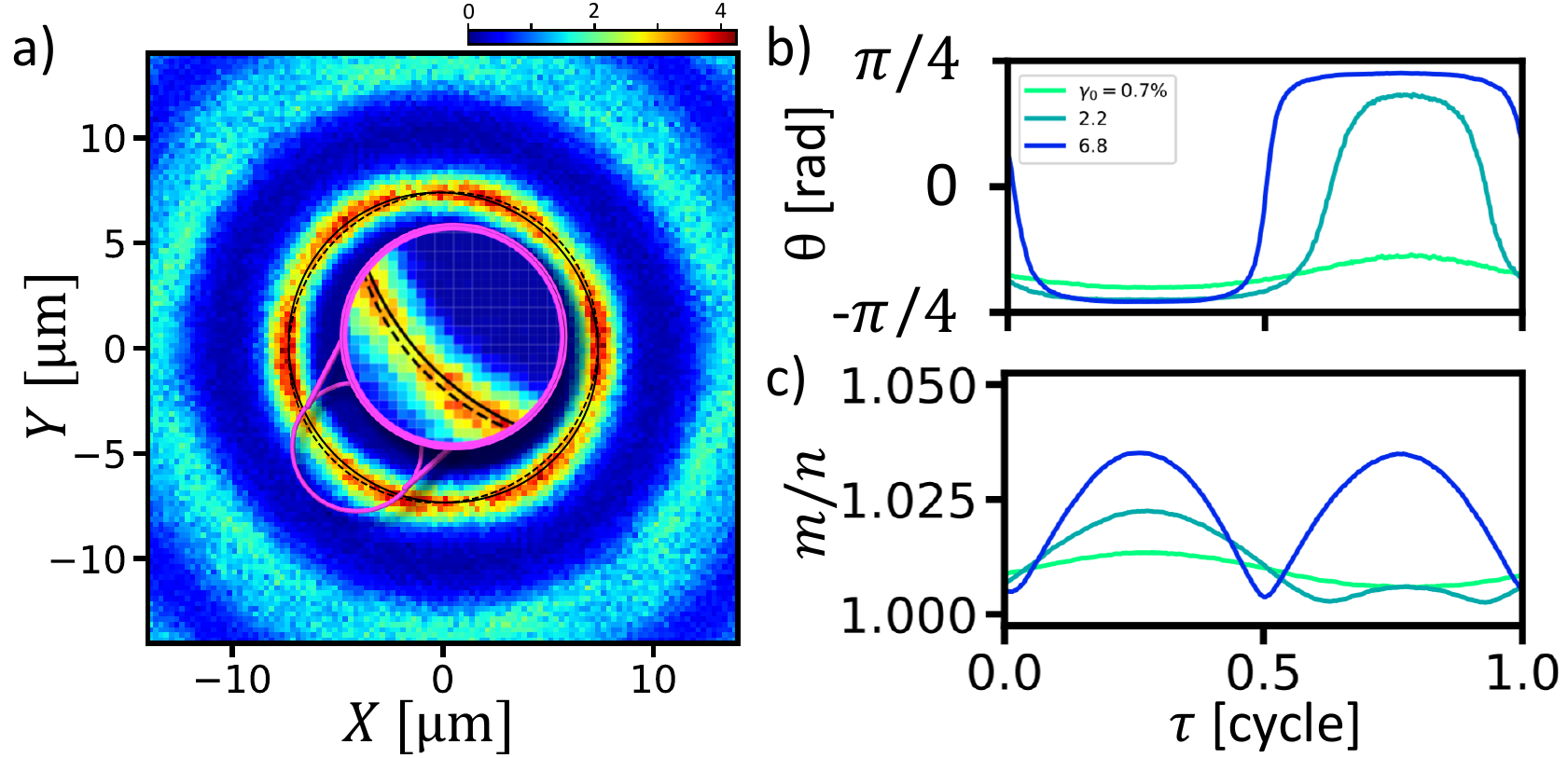}}
\end{figure*}

Next, we examine structural disorder, and its variation as a function of applied shear. The angle-dependent radial distribution function, $g(x,y)$, quantifies microstructural order \cite{larson99,deni15} (Fig.~\ref{fig:2}a). Crucially, a nearest-neighbor ring is observable in disordered systems composed of interacting particles \cite{verm05,parsi87}. In our experiments this ring deforms throughout shear (Fig.~\ref{fig:2}a and supplementary video), in agreement with previous observations \cite{parsi87,cheng11,seth11,ding15,maes17,deni15}. Throughout shear, the central ring is ellipsoidal. We can readily track the orientation and elongation of the ellipse throughout the shear cycle (Fig.~\ref{fig:2}b\&c); ellipse orientation and elongation provide a measure of the sample anisotropy. Far above yield, as the material is sheared in one direction and then the other, the microstructural anisotropy switches between two principal strain axes (oriented at $\pi/4 [rad]$ and $-\pi/4[rad]$, counter-clockwise from horizontal in Fig. \ref{fig:2}a); in this situation, microstructural anisotropy is responsive to the direction of imposed shear (Fig.~\ref{fig:2}b). Below yield, however, the microstructural anisotropy remains in its original orientation; shearing is not sufficient to overcome initial `frozen in' material structure. This phenomenon is apparent from changes in ring elongation (Fig.~\ref{fig:2}c) during the shear cycle. Note that above yield the microstructure elongates twice every shear cycle, at frequency $2\omega$, but below yield, the microstructure elongates only once per cycle at $\omega$. 

Microstructural anisotropy reveals a memory of the last direction the material was sheared above yield (Fig.~\ref{fig:2}). To remove internal stresses, each of our experiments is pre-sheared well above yield ($\gamma_0 \sim 50 \%$); nevertheless, this protocol imprints an anisotropy into the sample set by the last shear direction. Previously it was shown that this type of material memory is imprinted into $g(x,y)$ \cite{parsi87,keim13a,paul14}. Here, we find that this memory imprint is associated with the principal directions of shear (Fig.~\ref{fig:2}). Once a memory is stored, the memory is retained as long as the material is sheared elastically. Precisely when the material yields, all memory is lost, and the microstructure freely switches between both orientations. Taken together, these results indicate that materials store and express memories in the elastic regime but lose them in the plastic regime. Furthermore, recently we showed that orientational memory is stored most strongly within crystalline domains wherein particle rearrangements are most intensely suppressed \cite{teich20}. 

\begin{figure}
\centering
\includegraphics[scale=1]{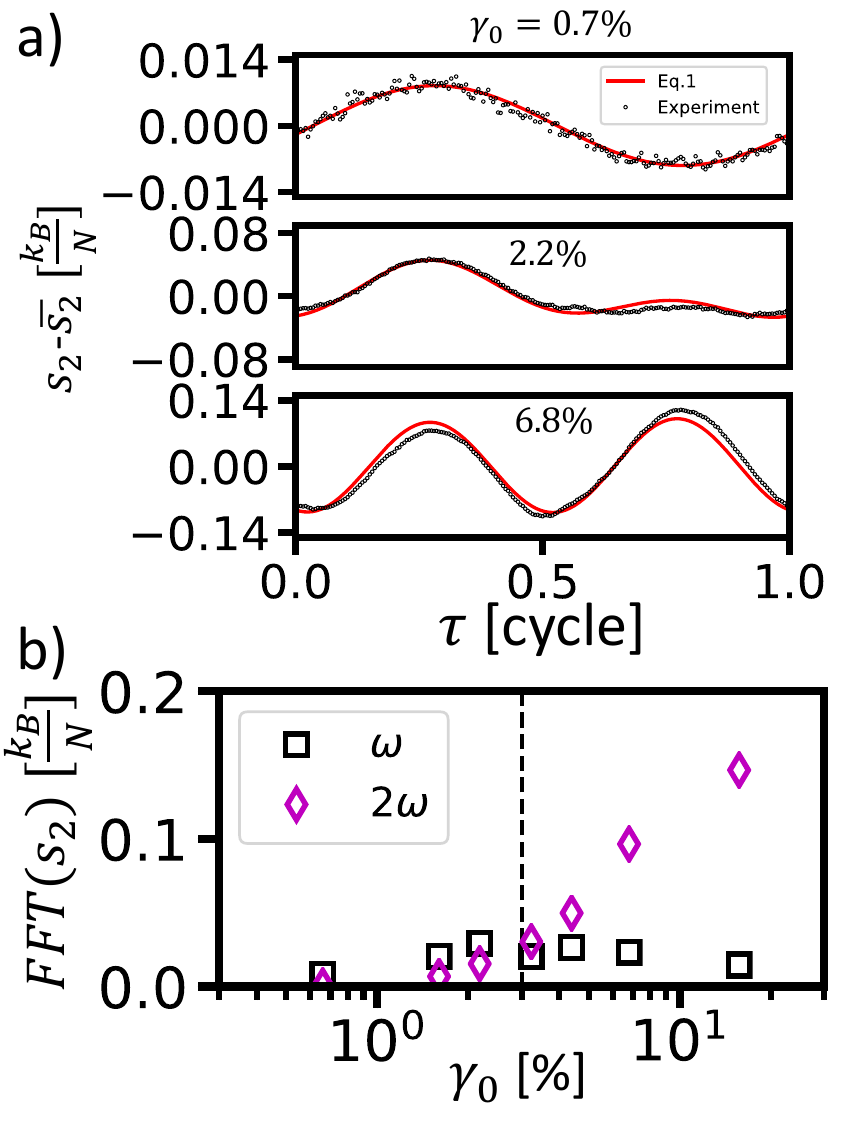}
\caption{ \textbf{ Entropy and material memories.} \small{Variation of entropy provides means for predicting system response to a given strain amplitude. (a) Excess entropy, with the mean value subtracted, follows a sinusoidal response. Below yield, its oscillation frequency is the shear cycle frequency. At yield, the excess entropy signal has components at both the driving frequency and twice the driving frequency: the material is beginning to forget its initial state. Above yield, the entropy response oscillates almost exclusively at twice the shear cycle frequency. Black dots indicate experimental data. Red lines are fits to equation~\ref{balance} with T as the only fitting parameter. The experimental data are averaged stroboscopically over 25 cycles. (b) Amplitudes associated with the first and second harmonics are present within the $s_2$ signals. Note, that the second and first harmonic amplitudes cross each other at the yield point, $\gamma_0=3\%$, designated by the vertical dashed line (- - -). } }
\label{fig:3}
\end{figure}

We now use excess entropy to characterize and relate observations about imprinted memory to the system microstructure. Above yield, we find that structural response is independent of the direction of shear (Fig.~\ref{fig:3}a, $\gamma_0=6.8\%$); when the material is sheared in either direction, the excess entropy increases and decreases as the shear is reversed. Ostensibly, the material cannot sustain a memory above yield, because it is continually forced out of meta-stable states within the energy landscape. Near yield, however, the direction of shear has an effect on structural response (Fig.~\ref{fig:3}a, $\gamma_0=2.2\%$). Notice, $s_2$ does not increase as the material is sheared over the second half of a sinusoidal shear cycle. Finally, below yield, the direction of shear is important; shear in one direction produces an increase in excess entropy, and shear in the other direction produces a decrease (Fig.~\ref{fig:3}a, $\gamma_0=0.7\%$). 

\begin{figure*}
\caption{ \label{fig:4} \textbf{ Comparisons of imposed force, microstructural excess entropy, and bulk rheology}. \small{a) The imposed force amplitude, $F_0$, normalized by the elastic force capacity, $F^*$, is plotted versus the excess entropy harmonic ratio, $s_{2,h}$ (in both mono-disperse and bi-disperse experiments). A fit of the data suggests a parabolic relationship (p-value:3.14x$10^{-13}$, and $r^2$:0.989), corroborating equation~\ref{harmonics}. b) The increase in the ratio of loss and storage moduli, $(G^{\prime\prime}/G^{\prime})$ versus strain amplitude in both the mono-disperse and bi-disperse experiments (same legend for mono-disperse and bi-disperse experiments as panel a). Yield is signaled by the rapid increase in parameter values at about $0.03$ strain amplitude. Inset: data from simulations employing Hertzian and Lennard-Jones interaction potentials. In both cases, markers are measured values and lines are predictions of equation~\ref{rheology}. c) Left and right hand sides of equation~\ref{rheology}. Notably, all parameters are measured. The solid diagonal line (\textbf{---}, slope of 1.0) represents equation~\ref{rheology}. The slope of the best fit to the data is 0.981, p-value:4.43x$10^{-26}$, and $r^2$:0.944. } }  
{\includegraphics{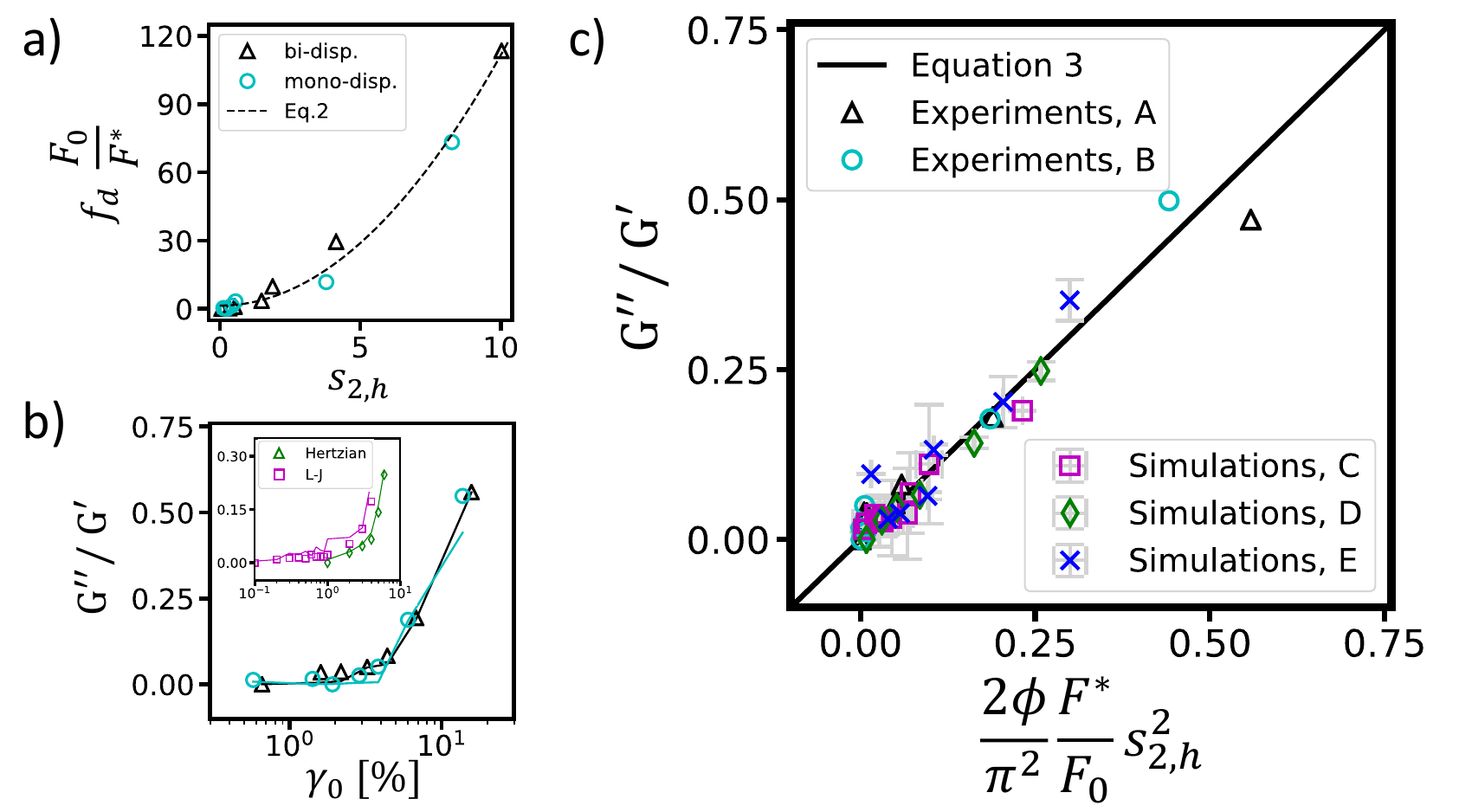}}
\end{figure*}

As seen in figure~\ref{fig:3}b, the $s_2$ signals are sinusoidal. The first harmonic ($\omega$) decays and the second harmonic ($2\omega$) grows with increasing strain amplitude. The first harmonic is dominant below yield, and the second is dominant above yield. Therefore, the amplitude of the first harmonic of $s_2$ provides quantification of a stored memory, and the amplitude of the second harmonic characterizes the degree to which memory of the initial state is lost. Notice, these first and second harmonic amplitudes cross each other near the yield point.  

To build a relationship between excess entropy and bulk rheology, we next investigate the connection of $s_2$ to the other dynamical metrics. For this comparison, we compute the ratio of the second to first harmonic amplitude, which we denote as $s_{2,h}$.  We can relate $s_{2,h}$ to several quantities in our system (Fig.~\ref{fig:4}). For example, $s_{2,h}$ scales with the product of $F^*/F_0$ and $f_d$ (Fig.~\ref{fig:4}a), where $F_0$ is the amplitude of the prescribed shear force. This relationship between dimensionless parameters suggests that when the imposed force on the system grows larger than $F^*$, the microstructure begins to permanently change, losing stored memory. Rapid variation of $f_d$ also signifies the transition. These findings build on recent work that links excess entropy and non-affine particle dynamics \cite{gallo20b,inge17}. Note that the scaling in the present case is quadratic because $f_d$ varies nearly linearly with the imposed force, $F_0$ (see Supplemental Materials). Finally, we find that the product of $s^2_{2,h}$ and $F_0/F^*$ scales linearly with $G^{\prime\prime}/G^{\prime}$ (Fig.~\ref{fig:4}c). The scaling factor for this linear relationship is $2\phi/\pi^2$; here $\phi=\pi N a^2 / A$ quantifies the particle spatial density, $a$ is the average nearest neighbor distance derived from the first peak of $g(r)$ (Fig.~\ref{fig:1}c: inset), and $A$ is the total area of the observed sample or simulation. 

The yield phenomenology shown in Fig. \ref{fig:4}c depends on four dimensionless parameters: $F_0/F^*$, $s_{2,h}$, $G^{\prime\prime}/G^{\prime}$, and the packing density $\phi$. The ratio $F_0/F^*$ characterizes the shear force exerted on the material relative to the force required to cause rearrangements; when $F_0/F^* \geq1$ plasticity is non-negligible. The microstructural quantity $s_{2,h}$ provides a metric for whether a material's response is dominated (or not) by memory as it experiences oscillatory strain; this microstructural property can be interpreted as the degree of plastic response. Finally, a familiar ratio quantifies the bulk rheological response of the material: ($G^{\prime\prime}/G^{\prime}$). All experimental (and simulation) data are collapsed using these dimensionless parameters, and a direct relationship between rheology, dynamics, and microstructure is experimentally established in the disordered solid. 

Numerical simulations complement the experiments. The simulations enable us to vary features of the disordered system that are difficult to control experimentally. In particular, we can test ideas regarding variation of inter-particle potential. Moreover, unlike the experimental system, which involves a fluid-fluid interface that gives rise to viscous drag on the particles, the simulations offer the possibility to study the validity of our new concepts in disordered materials without viscous drag. Thus, we have conducted shear simulations without viscous drag and with two different inter-particle interaction potentials: Lennard-Jones, a model for atomic glass, and Hertzian, a model for granular systems (see Methods). 

The simulations and experiments exhibit remarkably similar behaviors. Across both the experiments and simulations, a direct and common functional relationship between excess entropy and rheology is revealed (Fig.~\ref{fig:4}c). This relationship does not depend on the details of particle interactions, nor the amount of disorder. Further, since simulations do not involve a background fluid, the importance of hydrodynamic effects is ruled out. To test the limits of applicability of our numerical findings, we introduce varying amounts of Brownian motion into the Lennard-Jones simulations. At high thermal temperature, the particles rearrange due to Brownian motion in addition to shear stress, and memory cannot be formed. However, at low thermal temperature, the experimentally observed relationship between entropy and rheology holds (see Fig.~\ref{fig:4}b\&c). Moreover, we find that jamming is required for the storage of memories in both simulation systems. At low packing densities, where the system easily un-jams during shear, the relation is violated. The wide applicability of these ideas suggests the existence of a deeper theoretical formulation. Thus, in the remainder of this paper we outline how our results may be derived phenomenologically (for the full derivation see the Supplemental Materials). 

To elucidate the relationship between $s_2$ and the material properties ($G',G''$), we perform a simple energy balance. We start with the harmonic behavior in $s_2$. In this situation, energy is balanced in terms of accumulation, $T \Delta S_2$,  reversible (quasi-static) energy transfer, $ F^*x/2 $, and irreversible dissipation, $ f_d Fx$: 
\begin{equation} T \Delta S_2(t) = F^* x(t)/2 + f_d F(t) x(t). \label{balance} \end{equation} 
Here $x(t)$ is the displacement of the system boundary, $F(t)$ is the imposed shear force, and $T$ is a parameter (generally different from the thermal temperature) that converts differences in entropy to differences in energy \cite{bi15,ono02,bonn20,khab21}. Note that this equation would not apply in a system dominated by thermal motion, because we do not account for changes in entropy due to thermal fluctuations. The equation also implicitly reflects the requirement of jamming via the $F^*$ term. With a single fitting parameter, $T$, the changes in harmonic behavior in excess entropy are reproduced from below to above yield (Fig.~\ref{fig:3}a). 

The harmonic transition, associated with the excess entropy found in experiments and simulations, is captured by the first and second terms on the right-hand-side of equation \ref{balance}. $s^2_{2,h}$ is the ratio of those two terms: \begin{equation} s^2_{2,h}=f_d \frac{F_0}{F^*}  .\label{harmonics}\end{equation} 
This relation describes the harmonics data remarkably well (Fig.~\ref{fig:4}a). We next build on equation \ref{harmonics} by incorporating a finding of shear transformation zone theory, namely that elastic energy builds up in the microstructure until it is plastically released via non-affine rearrangement events \cite{falk11,falk98}. Quantitatively, this concept is represented as: $G^{\prime\prime} \propto N f_d G^{\prime}$, where $N$ is the number of total particles observed; when substituted into Eq.~\ref{harmonics} we obtain: \begin{equation} \frac{G^{\prime\prime}}{G^{\prime}} = \frac{2\phi}{\pi^2} \frac{F^*}{F_0} s^2_{2,h} .\label{rheology}\end{equation} 
Note, that each parameter in this expression is measured and is generally accessible in many systems. Across strain amplitudes, remarkable agreement is found between $G^{\prime\prime}/G^{\prime}$ measured in experiments and simulations, and the predictions by Eq.~\ref{rheology} (see Fig.~\ref{fig:4}b\&c).

\section{Conclusion} 

\begin{table*}
\small
  \label{tab:table1}
  \begin{tabular*}{1.0\textwidth}{@{\extracolsep{\fill}}lllllll}
    \hline
    ID & Type & Forces & Dispersity & Diameters & $\phi$ & $\Phi [\%]$ \\
    \hline
     A & Experiments & Dipole-dipole & Bi-disperse & 4.1, 5.6$\mu m$ & 14.02 & $\sim$31 \\
     B & Experiments & Dipole-dipole & Mono-disperse & 5.6$\mu m$ & 13.99 & $\sim$35 \\
     C & Simulations & Lennard-Jones & Bi-disperse & N/A & 5.03 & N/A \\
     D & Simulations & Hertzian & Bi-disperse & 0.84, 1.16 & 9.68 & 110 \\
     E & Simulations & Hertzian & Bi-disperse & 0.84, 1.16 & 10.12 & 120 \\
    \hline
  \end{tabular*}
  \caption{\ A summary of the properties of the systems presented, including variety of inter-particle force, particle dispersity, particle sizes, spatial density of particles, $\phi$, and simple area fractions of particles, $\Phi$. We note, particles are point particles in simulations, C; hence, diameters are not defined in system C. }
\end{table*}

Our results demonstrate that the yield transition of jammed systems has a configurational origin rooted in the persistence of material memory. We investigated the responses of several jammed systems undergoing cyclic shear deformation, incorporating aspects of STZ theory, excess entropy, and harmonic analysis into a single framework. The analysis reveals two new dimensionless parameters and three relations, derived phenomenologically, which connect particle configurations to bulk rheology. Importantly, the microstructural information needed, i.e., the radial distribution function, is available in myriad of scattering/microscopy experiments spanning length scales and particle types \cite{larson99}; thus, this analysis is accessible to experimentalists. In the future, it should be interesting to search for similar relations for other loading conditions, such as compression or steady shear, and to explore a wider array of particulate systems in which the particles are not simple spheres.

We have developed a framework to understand bulk properties of jammed materials under shear based on microstructural information. The findings hold potential to predict behavior of a broad range of dynamically arrested disordered materials including foams, gels, packings of nano- and micro-scale particles, and atomic/molecular glassy matter. Our findings, perhaps, also shed light on some deeper questions: in particular, the nature of entropy and the potential to use entropy ideas in far-from-equilibrium media. While entropy formulations for non-thermal systems have found utility in modeling disparate phenomena \cite{shann48,per16,jacob95}, its physical interpretation often remains mysterious. Disordered particulate packings appear to be particularly useful for clarifying this phenomenology, since their material structure can be interrogated with relatively simple methods.

\section{Methods}

\subsection{Experiments}

Using a custom built interfacial stress rheometer (ISR, SI Fig.~\ref{fig:ISR}), we simultaneously measure storage and loss moduli and track particle positions in 2D dense suspensions of athermal, repulsive particles. The ISR measures rheology by imposing force on a magnetic needle adsorbed at an interface between oil and water\cite{shah86}. A stationary wall is opposite the needle, so that shear is imposed over a distance visible by a microscope. The displacement of the rod is measured precisely with the microscope. With displacement (strain) and imposed force (stress), the storage and loss moduli are calculated \cite{brooks99,reyn08}. Additionally, the microscope is used to image the particles ($\sim 40,000$, from wall to needle) adsorbed at the interface. The particles include charges on their surfaces, so they exert dipole-dipole repulsive forces on each other \cite{avey00,mass10,park10}. At the particle densities in these experiments, these forces result in particle jamming, which we define as full kinematic restraint on each particle by its neighbors. In all data reported here the systems are in a sinusoidal, steady state. In the experiments, steady state occurs after five shear cycles. Twenty-five steady state cycles are used for calculations. For more information about these experiments and the calculations of $D^2_{min}$ see Refs.\cite{keim13b,keim14,keim15}.

An accessible quantity in our experiments is the two-body approximation of excess entropy, the difference between the system's entropy  and the entropy of an ideal gas in an equivalent state ($s_2 \sim s_{sys.}-s_{I.G.}$). Conveniently, this quantity is calculated from the radial distribution function, which is available in a wide range of experiments \cite{larson99}. The previously derived \cite{bara89} formula for excess entropy is:   
\begin{equation} \label{excess_entropy}
s_2 = - \pi \rho \int_{0}^{\infty} \big\{ g(r) ln [ g(r) ] - [g(r) -1] \big\} r dr
\end{equation}
where $\rho$ is the particle number density. We implement equation~\ref{excess_entropy} for each image in our experiments individually to collectively construct an entropy time signal, $s_2(t)$. For specifics of our excess entropy calculations, see Ref. \cite{gallo20b}. 

The network force, $F^*$ introduced in the paper is calculated based on inter-particle forces within the average neighborhood of particles. To make this measurement we estimate the average number of nearest neighbors around a particle as: \begin{equation} \label{N_NN} Z(R_c) = 2 \pi \rho \int_{0}^{R_c} g(r) r dr \end{equation}where $R_c$ values are shown as the horizontal axis in Fig.~\ref{fig:1}c. We estimate experimental inter-particle forces based on potentials measured in experiments and molecular dynamics simulations reported in Ref. \cite{park10}. An account of our estimate is included in the Supplemental Materials.  

\subsection{Simulations}
The data points for samples C were obtained using LAMMPS\cite{plimpt93}. At each strain amplitude, 10 two-dimensional ensembles of 10,000 bi-disperse Lennard-Jones particles\cite{widom87,patin16} were subjected to sinusoidal shear under periodic boundary conditions at constant confining pressure. The period of shearing was $100\times$ that of the LJ time-scale of the particles. Prior to shearing, the samples were dynamically equilibrated at $1\%$ of the glass-transition temperature\cite{patin16}. During strain-controlled shearing LAMMPS' Nos\'{e}-Hoover thermostat was used to maintain the samples at approximately $1\%$ of the glass-transition temperature. After 40 cycles of shearing, the shear stress was output for another 40 cycles for later use in the calculations of the dynamic moduli. We find that similar calculations at $9\%$ of the glass-transition temperature begin to violate our assumption of negligible thermal energy. 

For simulation samples D and E, we use HOOMD-blue \cite{glas15,ander08} to impose cyclic strain on 10 particle configurations for each of six strain amplitudes ($1,2,3,4,5,6\%$) at constant confining pressure. Ensembles are composed of jammed states of 50:50 bidisperse mixtures of 10,000 Hertzian particles. Ensembles are initialized from a randomly uniform probability distribution at a packing fraction below jamming, and subsequently quenched under FIRE minimization \cite{bitz06} whilst increasing the packing fraction until the desired pressure is reached. We then run a triangle wave shear protocol, imposing a small strain step of $10^{-4}\%$ and minimizing under FIRE after each step, until a total of 40 cycles have been completed. We calculate dynamic moduli based on the dominant frequencies of the resulting triangle waves.

\section{Acknowledgements}
We thank Doug Durian, Andrea Liu, Rob Riggleman, Ido Regev, and S\`{e}bastien Kosgodagan Acharige for fruitful discussions. We especially thank Andrea Liu for the generous contribution of computational resources for our simulations. This work is partially funded by University of Pennsylvania's MRSEC NSF- DMR-1120901 and by ARO W911-NF-16-1-0290. Celia Reina further thanks NSF Career Award, CMMI-2047506.

\newpage
\color{white}
a
\color{black}
\newpage
\section{Supplemental materials}
\setcounter{figure}{0}
\setcounter{equation}{0}

\begin{figure}
\centering
\includegraphics[scale=1]{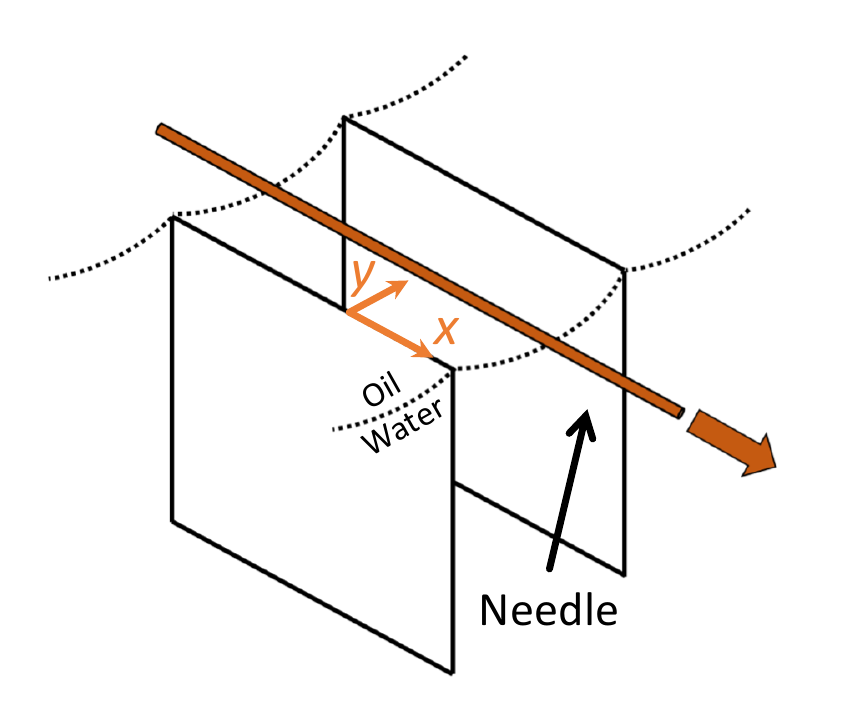}
\caption{ \textbf{Interfacial Stress Rheometer} \small{Sketch of interfacial stress rheometer (ISR). Dense colloidal monolayer sits on water-oil interface that is bounded by two upright parallel glass walls. A thin magnetic needle cyclically shears the monolayer using Helmholtz coils. Accurate rheometry is obtained by tracking needle position as a function of forcing; particle tracking is used to characterize material microstructure.} }
\label{fig:ISR}
\end{figure}

\subsection{Discussion of systems}

\begin{figure*}
\caption{ \label{fig:disorder} \textbf{Disorder increasing left to right} \small{ Crystalline regions visualized via sixfold bond orientation order, $\Psi_6$, measured from particle positions. Size of crystals decreases from left to right, indicating an increase in disorder. Colors help to indicate the lattice director (orientation) as a guide for the eye to help discern ordered and disordered domains. Dots with large size indicate $|\Psi_6| > 0.9$, and small dot size indicates $|\Psi_6| < 0.9$. (Scale bars: $100 \mu m$). a) Mono-disperse, dipole-dipole, experimental system B. b) Bi-disperse, dipole-dipole, experimental system A. c) Bi-disperse, Hertzian, simulation system D. d) Bi-disperse, Lennard-Jones, simulation system C. } }
{\includegraphics{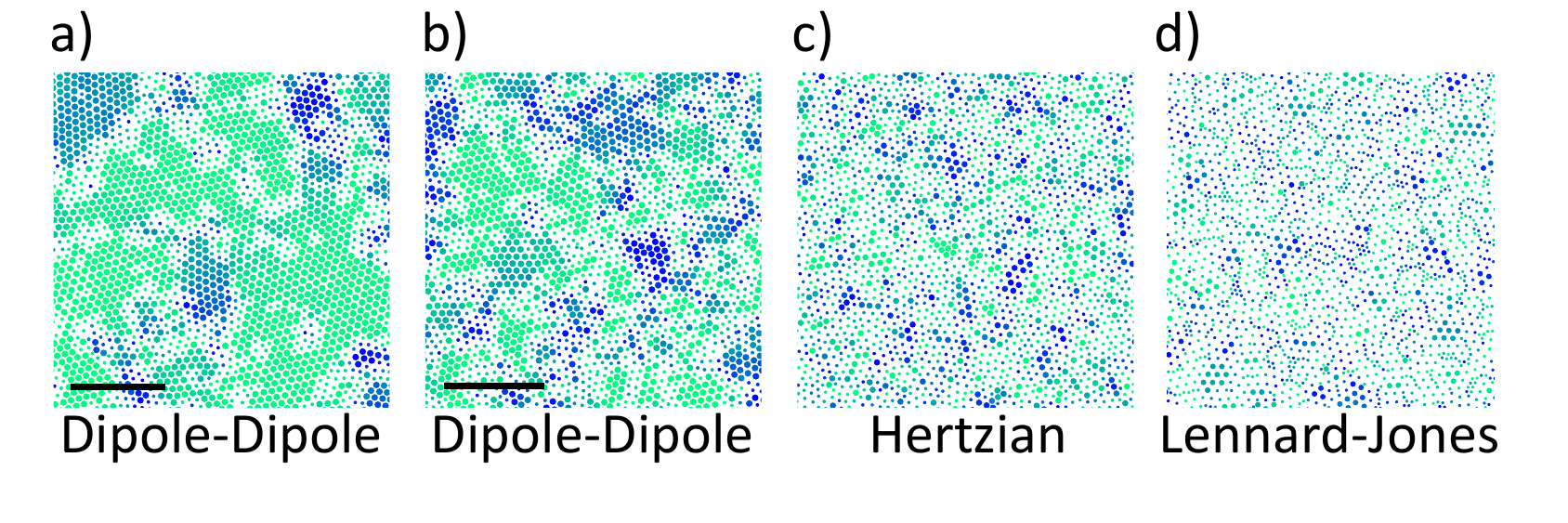}}
\end{figure*}

Here we make a few comments on the breadth of properties covered by our systems (as summarized in table~\ref{tab:table1}). First, we note that the systems studied span a large range of disorder, ranging from crystalline regions of several hundred particles (Fig.~\ref{fig:disorder}a) to merely a few (Fig.~\ref{fig:disorder}d). Second we note that, Lennard-Jones potentials are attractive at long distances. So our results hold for systems where some of the particles experience attraction, as long as the system is jammed. Finally, our systems span a wide range of length scales. Lennard-Jones systems are atomic scale. Dipole-dipole systems are colloidal scale. Hertzian systems are granular scale. One further difference among our systems is that our experiments include an intermediary fluid and an interface, whereas our simulations do not. This serves to explore the role of specific energy dissipation (viscous drag in the experiments) versus unspecific dissipation (the simulations).

\subsection{Scaling of imposed force and non-affine events}

\begin{figure}
\centering
\includegraphics[scale=1]{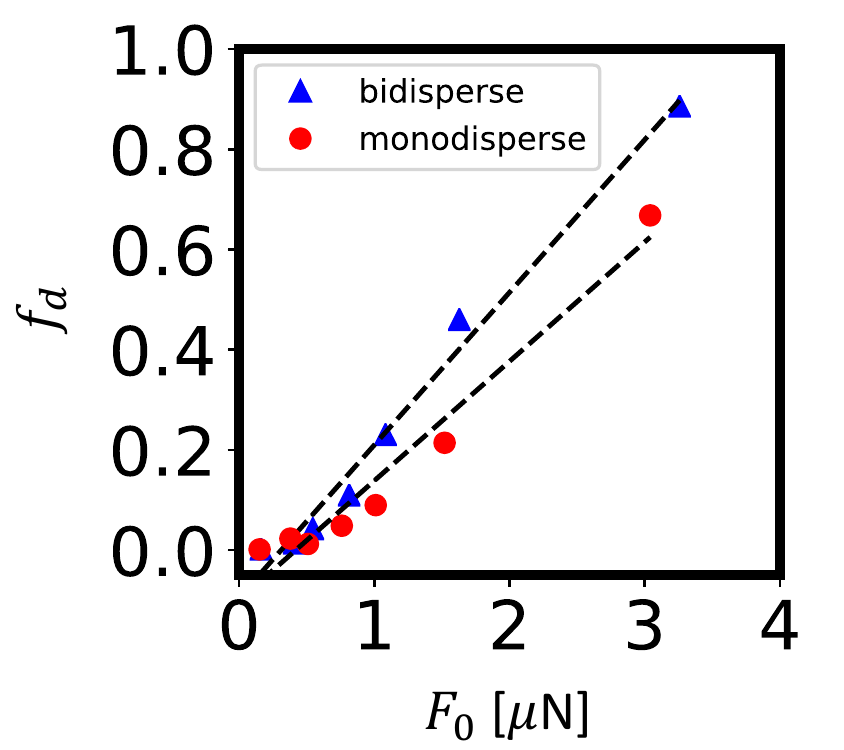}
\caption{ \textbf{Imposed force scales with fraction of dissipative events.} \small{ Within the linear rheology regime studied, the fraction of particles undergoing non-affine, disspative events scales linearly with the imposed force on the system. The dashed lines (- - -) are added to guide the eye. } }
\label{fig:fD2min_F}
\end{figure}

In the main text we identify that the entropic ratio $s_{2,h}$ varies as a quadratic in figure~\ref{fig:4}a. Our model, (main body equation~\ref{harmonics}), makes no prediction about the form of the scaling unless we know how force amplitude, $F_0$,  and the fraction of particles undergoing dissipative events, $f_d$, scale with each other. These values scale linearly with each other in our experiments (Fig.~\ref{fig:fD2min_F}), which results in a quadratic scaling. This is an approximation, as at very high strain amplitudes it is expected that the fraction of non-affine events will plateau at unity. But within the linear rheological regime studied here, this limit is not reached.

\subsection{ Experimental potentials}

In this section we describe our method of estimating the mean inter-particle potential of our experimental systems A and B. Sulfate latex spheres of $D_l = 5.6 \mu m$ and $D_s = 4.1 \mu m$ are adsorbed at an interface of decane and water. The sulfate latex groups cover the surfaces of the particles, providing a charge. The charges and the presence of the interface cause the particles to experience dipole-dipole repulsion with each other. The dipole-dipole form is: \begin{equation} \frac{u(r)}{k_BT}=a_2 \frac{1}{r^3} \label{potential} \end{equation} where $u$ is the potential, $k_B$ is Boltzmann's constant, $T$ is the thermal temperature, $a_2$ is the scaling constant, and $r$ is the center to center distance of the particles. In our bi-disperse system, the average separation between small particles is $r_{s s} = 7.53 \mu m$. Separation between large particles is $r_{ll} = 8.74 \mu m$.

This system is often used to study interfacial colloids; Park et al. \cite{park10} published a study that precisely measures the form of the interparticle potential quantitatively using Monte Carlo methods and optical tweezers. We used particles from the same manufacturer (Invitrogen Corporation, Carlsbad, CA) as Park et al. and followed the same particle cleaning procedure. They report that for particles of size $D_P = 3.1 \mu m$, the mean value of $\langle a_{2,P} \rangle = 5.1 \pm 2.4 \times 10^{-13} m^3$, where $P$ subscripts indicate Park et al.'s values. 

Within our bi-disperse system, the osmotic pressure is the same between large-large and small-small particles. Here the osmotic pressure is $-\frac{d^2u(r)}{dr^2}$. This allows us to write: \begin{equation}  \frac{\langle a_{2,s}\rangle}{r_{ss}^5}=\frac{\langle a_{2,l} \rangle}{r_{ll}^5}=\frac{\langle a_{2,p} \rangle}{r_{pp}^5} \label{osm_pressure} \end{equation} These equations are not linearly independent, so we extrapolate from our diameter-separation information ($[D_s,r_{ss}]$ and $[D_l,r_{ll}]$) to determine $r_{pp}$ using $D_P$. We find $r_{pp} = 6.72 \mu m$, $\langle a_{2,s}\rangle = 9.0 \times 10^{-13} m^3$, and $\langle a_{2,l}\rangle = 1.8 \times 10^{-12} m^3$. 

For the main body of the text we use the average weighted by particle numbers of $\langle a_{2,s}\rangle$ and $\langle a_{2,l}\rangle$ for the entire suspension as $\langle a_2 \rangle$. Forces are calculated as $F_{el.}(r) = -\frac{du}{dr}=\frac{3 \langle a_2 \rangle k_B T}{r^4}$.

\subsection{Phenomenological Derivation}

Briefly, as discussed in the main text, equation~\ref{balance} quantifies energy accumulated in the system, $TdS$ on the left hand side and the contributions from reversibly transferred and dissipated energy on the right hand side. Our experiments and simulations indicate that: \begin{equation} T \Delta S = F^* x/2 + f_{d} F x. \label{shear} \end{equation}
Here $T$ is a fit constant that converts changes in entropy to changes in energy. $S$ is the entropy of the entire system. $F^*$ is a property of the material that quantifies how jammed the particles are via the total force experienced by a typical particle with its neighbors. It is calculated as: $F^* = \rho \int{\int{ (-\frac{\partial u}{\partial r}) g(x,y) }} dxdy$, where $\rho$ is the number density of particles, $u$ is the inter-particle potential, and g(x,y) is the radial distribution function. Here $f_{d}$ is the fraction of particles undergoing dissipative events detected via non-affine rearrangements. Non-affine events are detected via $D^2_{min}$. See Refs \cite{keim13b,keim14,keim15} for details on this calculation. Specifically, $f_{d}=N_d/N$, where $N_d$ is the number of particles experiencing non-affine events and $N$ is the number of total particles observed. The proscribed shear force and resultant displacement of the shearing surface are $F$ and $x$ respectively. We define the $\Delta$ operation as the difference between entropy at time $t$ and the average entropy over an entire cycle of shear: $\Delta S = S(t)-\overline{S(t)}$.

We next summarize the specifics of our systems: most notably oscillatory shear and excess entropy. To apply equation \ref{shear} to the oscillatory shear cases considered in this paper, substitute in the time signals for shear surface displacement ($x(t)=x_0 sin(\omega t + \delta)$) and force ($F(t)=F_0 sin(\omega t)$) on the right-hand-side. On the left-hand-side, multiply by $Nk_B/Nk_B$: 
\begin{equation}
\begin{aligned}
Nk_BT \Delta (\frac{S}{Nk_B}) = \\
&\frac{F^* x_0}{2} sin(\omega t+\delta) \\
    &+ \frac{N_d}{N} F_0 sin(\omega t) x_0 sin(\omega t+\delta) \\
\end{aligned}
\end{equation}where $\omega$ is the frequency of the imposed force and $\delta$ is the resulting time lag between the imposed force and the resulting displacement. $\delta$ is an important physical parameter in rheology; it helps us to distinguish between solids, fluids, and everything in between. A fully elastic material has a $\delta=0 [rad]$; stress and strain are in phase as is seen from Hooke’s law. A fully viscous material has a $\delta= \pi/2 [rad]$; stress and strain are fully out of phase as is seen from Newton’s law of viscosity \cite{larson99}.

In our experiments, changes in pressure are negligible. Therefore, changes in absolute entropy are approximately the same as those for excess entropy ($ds_2 \sim ds-ds_{I.G.} \sim ds_{total}$); the ideal gas entropy is not expected to change. Notice, entropy has changed to lower case 's' to represent quantities that are normalized by $N$ and in units of $k_B$, which is convention. In simulations, entropy harmonics are directly calculated on $TS=E+PV$ because pressure, $P$, volume, $V$, and energy, $E$, are accessible \cite{ono02,scior99,bonn20}. Additionally, here we implement the product-to-sum trigonometric identity ($sin(u)sin(v) = (1/2)[cos(u-v)-cos(u+v)]$). Reorganizing gives: 
\begin{equation} 
\begin{aligned} 
\Delta s_{2} = \\
& \frac{F^* x_0}{2N k_B T} sin(\omega t+\delta) \\
& + \frac{N_d F_0 x_0}{2N^2 k_B T} \{ cos(\delta) - cos(2\omega t+\delta)\}. \\ 
\end{aligned}
\label{osc_shear}
\end{equation}Equation \ref{osc_shear} describes the evolution of a jammed system as it undergoes oscillatory shear and is fully non-dimensional. It is now apparent that the second term on the right-hand side (with $N_d$) has the second harmonic of the forcing frequency $2\omega$; this relation reproduces the frequency shift of the entropy signals in our simulations and experiments (main body Fig.~\ref{fig:3}a). The appearance of the second harmonic in the entropy signal captures well the transition to plasticity.

We investigate the yield transition further by taking the ratio of the first and second harmonics within frequency domain of $s_2$, ($s_{2,h}\equiv\frac{FFT_{s_2}(2\omega)}{FFT_{s_2}(\omega)}$), which follows from equation \ref{fig:4} as: \begin{equation} s^2_{2,h} = \frac{N_d}{N} \frac{F_0}{F^*}. \label{Harmonics} \end{equation}Equation \ref{Harmonics} is visualized in figure~\ref{fig:4}a of the main text. This scaling is quadratic because $N_d$ and $F_0$ scale linearly with each other (Fig.~\ref{fig:fD2min_F}). The square of $s_{2,h}$ in equation \ref{Harmonics}, is included so that linear relationships are retained throughout. 

From here we revisit an idea posited by Falk and Langer (Ref.\cite{falk98}): relaxation events are due to a local buildups of elastic energy that suddenly release (i.e. $G^{\prime\prime}\propto N_d G^{\prime}$). Recently quantified for above yield cases in Ref.\cite{keim14} and here expanded to below yield, $G^{\prime\prime} = \frac{2a^2}{\pi A} N_d G^{\prime}$, where $a$ is the first peak distance of g(r) and $A$ is the area of observation. Substituting this equation into equation~\ref{Harmonics} for $N_d$ gives: \begin{equation} \frac{G^{\prime\prime}}{G^{\prime}} = \frac{2Na^2}{\pi A} \frac{F^*}{F_0} s^2_{2,h} = \frac{2\phi}{\pi^2} \frac{F^*}{F_0} s^2_{2,h} \label{Rheology} \end{equation}which allows us to relate the bulk material response directly to measurable microstructural properties without the use of fitting parameters. Equation~\ref{Rheology} is visualized in Fig.~\ref{fig:4}b\&c. $\phi$ quantifies particle density as $\phi=\pi N a^2 / A$, which implicitly takes $a$ as an effective particle diameter. This relation reveals that the yielding transition of jammed materials is specified by four dimensionless groups based on imposed force, particle density, a memory based dimensionless entropy, and the bulk response.

\bibliography{references.bib}
\bibliographystyle{rsc}

\end{document}